\documentclass[prl,twocolumn,showpacs,amsmath,amssymb]{revtex4}
\usepackage{graphicx}

\newcommand{\ket}[1]{\, | #1 \rangle}

\newcommand{\de}{\delta}

\newcommand{\ve}{\varepsilon}
\newcommand{\sgn}{\operatorname{sgn}}

\newcommand{\bk}{b^{\dagger}}

\newcommand{\hn}{\hat{n}}

\newcommand{\be}{\begin{equation}}
\newcommand{\ee}{\end{equation}}
\newcommand{\bea}{\begin{eqnarray}}
\newcommand{\eea}{\end{eqnarray}}
\newcommand{\besa}{\begin{subequations}\begin{eqnarray}}
\newcommand{\eesa}{\end{eqnarray}\end{subequations}}

\begin{document}

\title{Scattering resonances and two-particle bound states  
of the extended Hubbard model}

\author{Manuel Valiente}
\author{David Petrosyan}
\affiliation{Institute of Electronic Structure \& Laser, FORTH,
71110 Heraklion, Crete, Greece}

\date{\today}

\begin{abstract}
We present a complete derivation of two-particle states of the
one-dimensional extended Bose-Hubbard model involving attractive 
or repulsive on-site and nearest-neighbour interactions. We find
that this system possesses scattering resonances and two families
of energy-dependent interaction-bound states which are not present 
in the Hubbard model with the on-site interaction alone. 
\end{abstract}


\maketitle

%%%%%%%%%%%%%%%%%%INTRODUCTION%%%%%%%%%%%%%%%%%%%%%%%

Among the tight-binding lattice models of condensed matter
physics, the Hubbard model plays a fundamental role \cite{SolStPh}. 
In its basic form, the Hubbard model describes particle tunneling 
between adjacent lattice sites as well as short range (contact) interaction 
between the particles on the same lattice site. Despite apparent simplicity,
this model is very rich in significance and implications for the many
body physics on a lattice \cite{bosMI}. This is perhaps most profoundly 
manifested with numerous important experimental and theoretical 
achievements with cold neutral atoms trapped in optical 
lattice potentials \cite{optlattMI,OptLatRev}, wherein the Hubbard 
model is being realized with unprecedented accuracy. 

The next level of generalization pertaining to, e.g., electrons in 
a crystal \cite{MRRrev} or dipolar atoms \cite{DipAt} or molecules 
\cite{DipMol} in an optical lattice, yields the extended
Hubbard model involving longer-range interactions between the 
particles on the neighboring lattice sites. Under certain conditions,
namely, for hard-core bosons or strongly interacting fermions at 
half-filling,  the extended Hubbard model can be mapped onto various 
lattice spin models \cite{Sadchev}. 

A remarkable Hubbard model effect demonstrated in a seminal experiment
of Winkler {\it et al.} \cite{KWEtALPZ} is the binding of pairs of 
particles into composite objects by the on-site interaction, which 
can be either repulsive or attractive \cite{MVDP2,molmer}. 
In the present paper, we study the one-dimensional extended 
Bose-Hubbard model involving attractive or repulsive on-site 
and nearest-neighbour interactions. We present a complete 
derivation of two-particle states of the system and show that it 
possesses scattering resonances and energy-dependent interaction-bound
states which are not present in the Hubbard model with the on-site 
interaction alone.

%%%%%%%%%%%%%%%%%%%FORMULATION%%%%%%%%%%%%%%%%%%%%%%%

The Hamiltonian of the extended Hubbard model reads
\bea
H &=& \sum_{j} \ve_j \hat{n}_j - J \sum_{j} (\bk_j b_{j+1} + \bk_{j+1} b_j ) 
\nonumber \\ & &
+ \frac{U}{2} \sum_{j} \hat{n}_j(\hat{n}_j-1) 
+ V \sum_{j} \hat{n}_j\hat{n}_{j+1} , \label{BHHam}
\eea
where $\bk_{j}$ ($b_{j}$) is the creation (annihilation) operator 
and $\hn_j = \bk_{j} b_{j}$ the number operator for a boson at $j$th 
lattice site with energy $\ve_j$, $J$($>0$) is the tunnel coupling 
between adjacent sites, and $U$ and $V$ are, respectively, the on-site
and nearest-neighbor interactions which can be attractive or repulsive. 

We seek two-particle eigenstates of Hamiltonian (\ref{BHHam}) in
a homogeneous lattice, $\ve_j = \ve$ for all $j$. For convenience,
we set $\ve = 0$, which amounts to a shift of the zero of energy.  
We then expand the state vector in terms of the non-symmetrized 
coordinate basis $\{\ket{x_j , y_{j'}}\}$ as
$\ket{\Psi} = \sum_{j,j'} \Psi(x_j,y_{j'}) \ket{x_j , y_{j'}}$,
where $x_j\equiv d j$ and $y_{j'} \equiv d j'$ are the particle positions 
and $d$ is the lattice constant. The standard (symmetrized) bosonic basis 
is recovered via the transformation $\ket{2_j} = \ket{x_j, y_{j}}$ and 
$\ket{1_j,1_{j'}} = \frac{1}{\sqrt{2}} (\ket{x_j, y_{j'}} + \ket{y_j, x_{j'}})$
($j\neq j'$). Defining the center of mass $R = \frac{1}{2}(x + y)$ and 
relative $r = x - y$ coordinates, the two-particle wavefunction 
can be factorized as $\Psi(x,y) = e^{i K R} \, \psi_K (r)$, 
where the relative coordinate wavefunction $\psi_K (r)$ depends 
on the center-of-mass quasi-momentum $K \in [-\pi/d,\pi/d]$ as a 
continuous parameter. The eigenvalue problem  
$H \ket{\Psi}=E \ket{\Psi}$ then reduces to the three-term difference 
equation
\bea
&& -J_K\big[ \psi_K (r_{i-1}) + \psi_K (r_{i+1}) \big] 
\nonumber \\ && 
+ \big[U \de_{r,0} +V (\de_{r,d}+\de_{r,-d})- E_K \big] \psi_K (r_i)
= 0  ,   \label{tpwfr}
\eea
with $ J_K \equiv 2 J \cos(Kd/2)$ and $r_i = d i$ ($i = j - j'$).
The above equation admits two kinds of solutions, corresponding 
to the scattering states of asymptotically-free particles and 
to the interaction-bound states of particle pairs.

%%%%%%%%%%%%%%%Scattering states%%%%%%%%%%%%%%%%%%

%%%%%%%%%%%%%%%%%%%%%%%%%%%%%%%%%
\begin{figure}[t]
\includegraphics[width=0.42\textwidth]{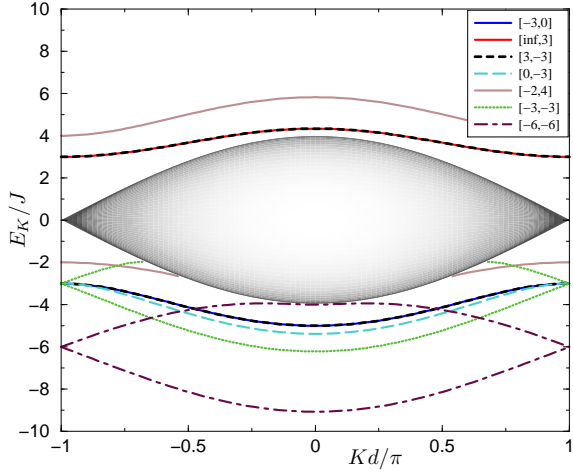}
\caption{Energies versus the center-of-mass momentum $K$
for a pair of bosons in a 1D lattice described by the extended 
Hubbard model. The continuum spectrum corresponds to energies
$E_{K,k}$ of the scattering states, with the shading proportional
to the density of states $\rho(E,K)$. The lines correspond to
energies $E_{K}$ of the two particle bound states for various 
values of interaction strengths $[U/J,V/J]$.}
\label{fig:spectrum}
\end{figure}
%%%%%%%%%%%%%%%%%%%%%%%%%%%%%%%%%

We begin with the analysis of scattering solutions of Eq.~(\ref{tpwfr}). 
They are most straightforwardly obtained with the standard symmetrized ansatz 
$\psi_{K,k}(r_i\neq 0) \propto e^{-ik|r_i|} + e^{2 i \delta_{K,k}}  e^{ik|r_i|}$
representing plane waves undergoing a scattering phase shift $\de_{K,k}$. 
This immediately yields the eigenenergies $E_{K,k}= -2J_K \cos (kd)$, which 
are equal to the sum of Bloch bands $\epsilon_{x,y} = -2 J\cos[(K/2 \pm k)d]$
of two (asymptotically) free particles $x,y$ with relative quasimomentum $k$ 
\cite{molmer,MVDP2}. For a given value of the center-of-mass momentum $K$, 
and thereby $J_K$, the lowest $E_{K,0} =  -2 J_K$ and highest 
$E_{K,\pi} =  2 J_K$ energy states are attained, respectively, 
at $k \to 0$ and $k \to \pi/d$. 
The continuum of energies $E_{K,k}$ and the corresponding density of states 
$\rho(E,K) \propto \partial (k d)/ \partial E = [ (2 J_K )^2 - E^2 ]^{-1/2}$
are shown in Fig.~\ref{fig:spectrum}. The wavefunction of scattering 
states is given by 
\besa
&& \psi_{K,k}(0) = \cos (\de_{K,k}^{(0)}) 
\frac{ \cos (kd + \de_{K,k}) }{ \cos (kd+\de^{(0)}_{K,k})} , \\
&& \psi_{K,k} (r_i\neq 0) = \cos (k |r_i| + \de_{K,k}) , \label{psiscat}
\eesa
where the phase shifts $\de^{(0)}_{K,k}$ and $\de_{K,k}$ are defined through
\besa
\tan (\de^{(0)}_{K,k}) &=& - \frac{U }{2 J_K \sin(k d)}  , \\
\tan (\de_{K,k}) %\nonumber \\ && \quad
&=& \frac{J_K U +  [2 J_K \cos (kd) + U ] \, V \cos (kd)}
{ \{ U V - 2 J_K [J_K-V\cos{(kd)} ] \} \sin(k d) } . \qquad  \label{tdelta}
\eesa

%%%%%%%%%%%%%%%%%%%%%%%%%%%%%%%%%
\begin{figure}[t]
\includegraphics[width=0.48\textwidth]{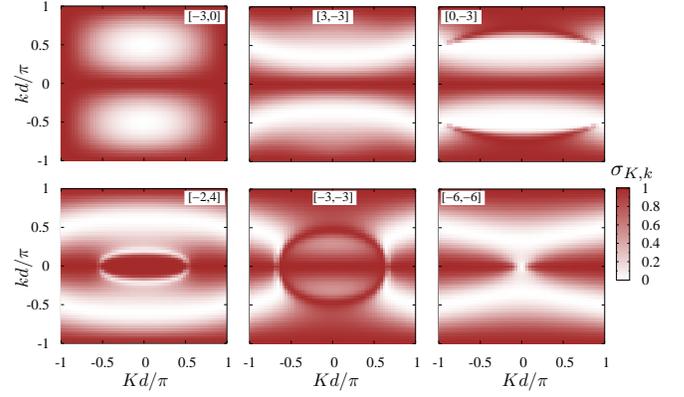}
\caption{Two-particle scattering cross-section $\sigma_{K,k}$ 
versus the center-of-mass $K$ and relative $k$ quasi-momenta 
for several values of interaction strengths $[U/J,V/J]$.
Flipping simultaneously the sign of both $U$ and $V$
is equivalent to shifting $k \to k + \pi/d$.}
\label{fig:sigmas}
\end{figure}
%%%%%%%%%%%%%%%%%%%%%%%%%%%%%%%%%

Note that when $V=0$, we have $\de_{K,k}= \de^{(0)}_{K,k}$, 
and the above expressions reduce to those of \cite{MVDP2} with 
all the consequences discussed there. Here we examine the 
role of nearest-neighbour interaction $V \neq 0$ due to which 
the scattering amplitude $f(\de_{K,k})= \frac{1}{2}(e^ {2i\de_{K,k}}-1)$
or, for that matter, the cross-section  
$\sigma_{K,k} = \left|f\left(\delta_{K,k}\right)\right|^2 =\sin^2 (\delta_{K,k})$ 
can exhibit novel intriguing features seen in Fig.~\ref{fig:sigmas}. 
The cross-section vanishes at $\de_{K,k} = 0$, which, according to 
Eq.~(\ref{tdelta}), requires 
$E_{K,k} = \frac{1}{2} U [1 \pm \sqrt{1 - 8 J_K^2 /(UV)}]$ 
or, more explicitly, 
$\cos(kd) = - U/(4 J_K) [ 1 \pm \sqrt{ 1 - 8 J_K^2 /(UV)} ]$,
with the condition $8 J_K^2/(UV) \leq 1$. 
The cross section attains its maximal value $\sigma_{K,k}=1$ at 
$\de_{K,k} = \pm \pi/2$, which corresponds to a pair of impenetrable
(hard-core) bosons. According to Eq.~(\ref{tdelta}), this happens when 
either $\sin(kd) \to 0$ and $J_K \neq \mp UV/(U+2V)$ (for $kd \to 0,\pi$, 
respectively), or $E_{K,k} = U - 2 J_K^2/V$, the last equality 
yielding the explicit condition $\cos(kd) = J_K/V - U/(2 J_K)$.

\begin{figure}[t]
\includegraphics[width=0.38\textwidth]{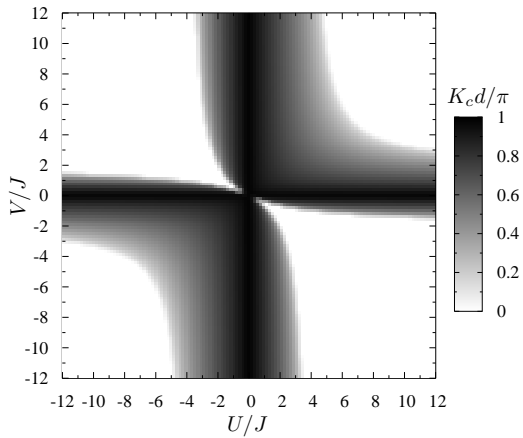}
\caption{The $U,V$ diagram for the existence of the second bound 
solution $|\alpha_K^{(2)}| < 1$ of Eq.~(\ref{cubalpha}) with 
$K_c < |K| \leq \pi / d$, with shading proportional to 
$K_c \in [0,\pi / d]$.}
\label{fig:UVdiag}
\end{figure}

At the edges of the scattering band, $k d \to 0,\pi$, we can define 
generalized 1D scattering lengths $a_K^{(0,\pi)}$ via \cite{olshanii,molmer,Naidon}
\be
a_K^{(0,\pi)} = - \!\! \lim_{k d \to \genfrac{}{}{0pt}{}{0}{ \pi }}
\frac{\partial \de_{K,k}}{\partial k}
= \frac{UV - 2 J_K (J_K \mp V)}{UV \pm J_K (U+2V)} \, d .
\ee
Thus, the scattering length $a_K$ vanishes at the bottom of 
scattering band, $k \to 0$, when $J_K = \frac{1}{2}[V \pm \sqrt{V(2U+V)}]$, 
and at the top of the band, $k \to \pi/d$, when 
$J_K = - \frac{1}{2}[V \pm \sqrt{V(2U+V)}]$, 
with the condition $V(2U+V) \geq 0$. The scattering length 
diverges when $J_K = \mp W$ for $k d \to 0,\pi$, where 
$W \equiv UV/(U+2V)$. In other words, the divergence of $a_K^{(0,\pi)}$ 
occurs when the center-of-mass quasi-momentum $|K|$ is equal to 
\be
K_R = \frac{2}{d} \arccos \left(\mp \frac{W}{2J} \right) , \label{KR} 
\ee  
with the condition $0 \leq \mp W/2J \leq 1$ for $k d \to 0,\pi$, respectively. 
As will become apparent from the proceeding discussion, $K_R$ indicates
the emergence of scattering resonances associated with the bound states 
(see Fig.~\ref{fig:UVdiag}).

%%%%%%%%%%%%%%%Bound states%%%%%%%%%%%%%%%%%%

We now consider the two-particle bound states. 
Using the exponential ansatz $\psi_K(r_i \neq 0) \propto \alpha_K^{|i|-1}$
in Eq.~(\ref{tpwfr}), after little algebra, we obtain
\be
J_K V \alpha_K^3+ (V U - J_K^2) \alpha_K^2 + J_K(V + U) \alpha_K + J_K^2 = 0 . 
\label{cubalpha} 
\ee
Solutions of Eq.~(\ref{cubalpha}) with $|\alpha_K| < 1$ determine 
the normalized bound states 
\besa
&& \psi_K(0) = \mathcal{N} \frac{2J_K \alpha_K}
{U \alpha_K + J_K (\alpha_K^2 +1)} \equiv \mathcal{N} \varphi_K , \\
&& \psi_K(r_i\neq 0) = \mathcal{N} \alpha_K^{|i|-1}, 
\eesa
where $\mathcal{N} \equiv [\varphi_K^2 +2/(1-\alpha_K^2)]^{-1/2}$,
with the energies given by 
\be
E_K = -J_K \frac{1+\alpha_K^2}{\alpha_K} . \label{bdEK} 
\ee
Note that complex solutions of Eq.~(\ref{cubalpha}) do not correspond to 
the bound states, even if $|\alpha_K|<1$, since the energy of Eq.~(\ref{bdEK})
should be real.

%%%%%%%%%%%%%%%%%%%%%%%%%%%%%%%%%
%%%%%%%%%%%%%%%%%%%%%%%%%%%%%%%%%

It can be shown that Eq.~(\ref{cubalpha}) admits at most two solutions
corresponding to the bound states. Obviously, for noninteracting 
particles $U = V =0$, there can be no bound state. 
For $U \neq 0$ and $V =0$ \cite{MVDP2} and for $U=0$ and $V\neq 0$, 
there is only one bound solution $\alpha_K^{(1)}$ at any $K$. 
For any other values of $U,V \neq 0$, the first bound solution
$\alpha_K^{(1)}$ exists at any $K$, and the second bound solution 
$\alpha_K^{(2)}$ exists at $|K| > K_c$, where the critical $K_c$ is
shown in Fig.~\ref{fig:UVdiag} and is defined as
\be
K_c = \begin{cases}
K_R & \mbox{if $|W/ 2J| \leq 1$} \\
0^-   & \mbox{otherwise}  
\end{cases} .
\ee
We thus see that the scattering length diverges when the second bound 
state approaches the edge of the scattering continuum. This signifies 
the appearance of the scattering resonance at $|K| = K_R$ which is 
determined by Eq.~(\ref{KR}) under the condition $|W/2J| \leq 1$.
When, however, this condition is not satisfied, no scattering resonance
is present, and the second bound state exists for all $K \in [-\pi/d,\pi/d]$
with the energy below or above the scattering continuum depending on 
whether $W \equiv  UV/(U+2V)$ is negative or positive, respectively
(see Fig.~\ref{fig:spectrum}).

In general, analytic expressions for the bound solutions of 
Eq.~(\ref{cubalpha}) are too cumbersome for detailed inspection, 
but several special cases yield simple instructive results.
(i)~With only the on-site interaction $U \neq 0$ and $V=0$,
there is one bound solution  
$\alpha_K^{(1)} = ( U - E_K ) /(2 J_K)$ with the corresponding energy 
$E_K = \sgn(U) \sqrt{U^2 + 4 J_K^2}$, as was discussed in \cite{MVDP2}.
(ii)~With very strong on-site interaction $|U| \to \infty$ and $V \neq 0$,
the first bound solution is trivial, $\alpha_K^{(1)} =0$ and $E_K = U$, 
representing an infinitely bound pair. The second more relevant solution
$\alpha_K^{(2)} = - J_K/V$ and $E_K = V + J_K^2/V$ describes a pair
of hard-core bosons, $\psi_K(0) = 0$, which are bound by the 
nearest-neighbour interaction $V$ provided $|J_K/V| < 1$ \cite{Scott}. 
Note that in this limit we have $W=V$, and the last condition for 
the existence of the second bound state again reduces to $|K| > K_c$.
(iii)~Finally, a rather curious and simple case is realised with $U =-V$. 
The first bound solution is similar to that in (i), but with $U$ replaced
by $V$, $\alpha_K^{(1)} = ( V - E_K ) /(2 J_K)$ and the energy 
$E_K = \sgn(V) \sqrt{V^2 + 4 J_K^2}$, which corresponds
to binding mainly by the off-site interaction. The second 
bound solution is similar to that in (ii), but now with $V$ replaced
by $U$, $\alpha_K^{(2)} = - J_K/U$ and $E_K = U + J_K^2/U$, provided 
$|J_K/U| < 1$ (note that now $W=U$). This solution corresponds to the 
on-site interaction binding. 

%%%%%%%%%%%%%%%%%%%%%%%%%%%%%%%%%
\begin{figure}[t]
\includegraphics[width=0.48\textwidth]{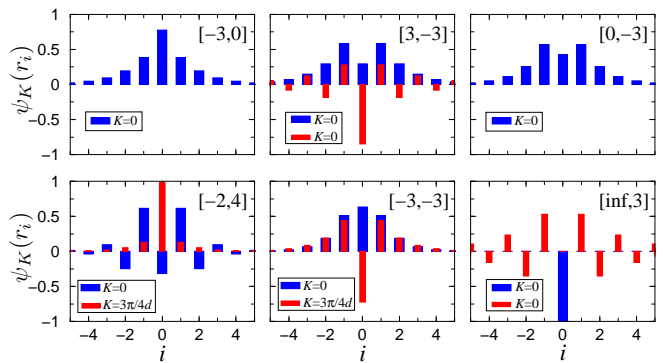}
\caption{Relative coordinate wavefunctions $\psi_K(r_i)$ of the 
two-particle bound states for several values of interaction strengths 
$[U/J,V/J]$. Thick (blue) bars correspond to the first bound solution
of Eq.~(\ref{cubalpha}) at $K=0$. Thinner (red) bars correspond 
to the second bound solution at $K=0$, if it exists for all $K$,
or at $K=3\pi/4d$, if it exists for $|K| > K_c$ 
(cf. Figs.~\ref{fig:spectrum} and \ref{fig:UVdiag}).}
\label{fig:bspsir}
\end{figure}
%%%%%%%%%%%%%%%%%%%%%%%%%%%%%%%%%

More generally, when the on-site $U$ and off-site $V$ interactions
have different sign, the first bound state is associated mainly with 
the stronger (in absolute value) interaction, and the second bound 
state with the weaker one. When, however, $U$ and $V$ have the same
sign and comparable strength, the bound states have mixed nature
in the sense that both interactions significantly contribute to the 
binding. To illustrate the foregoing discussion, in Fig.~\ref{fig:bspsir}
we show the wavefunctions of bound states for several cases pertaining
to the on-site, off-site and mixed binding, while the corresponding 
energy dispersion relations are plotted in Fig.~\ref{fig:spectrum}.

%%%%%%%%%Conclusions%%%%%%%%%%%%%%%

To conclude, we have derived a complete solution of the two-body problem
in a one-dimensional extended Hubbard model. We have found that depending
on the strength of the on-site and nearest-neighbour interactions, this
system possesses one or two families of bound states as well as scattering 
resonances corresponding to the degeneracy of the bound and scattering
states. Our results pertaining to the on- and off-site pairing mechanisms 
might be relevant to the studies of high-$T_c$ superconductivity \cite{MRRrev}
and to the more recent quest to realize related physics with cold 
trapped atoms or molecules in optical lattices \cite{EH-OL,PSAF}.

\begin{acknowledgments}
This work was supported by the EC Marie-Curie Research Training Network EMALI.
\end{acknowledgments}

%%%%%%%%%BIBLIOGRAPHY%%%%%%%%%%%%%%%

\end{document}